# Modeling and control of traveling wave structures with arbitrary beam loading

Zheqiao Geng[1,†], Riccardo Zennaro, Alessandro Citterio
[1]Paul Scherrer Institut, 5232 Villigen PSI, Switzerland

**ABSTRACT**: Traveling wave (TW) structures are widely used in linear or circular accelerators for beam acceleration. Unlike the standing wave cavities, the overall accelerating voltage cannot be measured by the probes installed in the TW structures. This creates difficulties measuring and regulating the accelerating voltage and phase felt by the beam when it passes through the structure. In this paper, a model of TW structures is developed, allowing us to estimate the accelerating voltage and phase from the measurement of the RF drive power input to the TW structure. The beam loading effects, i.e., the beam-induced accelerating voltage and phase in the TW structure, are also included in the model. With the TW structure model, control methods are proposed to stabilize the accelerating voltage and phase felt by the beam based on feedback, feedforward, and optimization. The validity of the TW structure model is confirmed with the test results at the SwissFEL with improved beam stability under the same RF feedback controller. This work provides a firm basis for developing the multiple-bunch operation strategy for SwissFEL required by the future upgrade with more beamlines.

## I. INTRODUCTION

Traveling wave (TW) structures are widely used in particle accelerators for beam acceleration, especially in normal conducting electron/positron linear accelerators [1-3]. In a TW structure, RF drive power establishes electromagnetic fields in the cells as a traveling wave. The phase velocity of the traveling wave is chosen to match the particles' flight speed to maintain continuous acceleration, and the group velocity is typically much lower than the particle speed to establish the required electric gradient in the structure.

In a standing wave cavity, probes can be installed to pick up the cavity field. The amplitude and phase of a cavity probe signal can fully describe the accelerating voltage and phase experienced by the beam after a proper calibration. However, in a TW structure, the beam experiences an integrated electric field along the beam path. To measure the total accelerating voltage and phase experienced by the beam, the electric fields at all positions on the beam path should be known, which requires installing many probes along the structure (better within every cell). This is not practical for TW structures because the reliability and field quality of the structure will be degraded significantly with more probes. Typically, we only install RF pickups in the waveguides upstream the structure's input coupler and downstream the structure's output coupler, i.e., at the load. This raises the question of how to measure the accelerating voltage and phase experienced by the beam just using the measurements of these two signals.

If the TW structure operates in pulse mode and only 1 bunch is accelerated within a pulse, we can average the waveform measured at the structure input over a time window to represent the field experienced by the beam [4]. The averaging window length should match the structure's filling time [5] with its position synchronized to the specific region interacting with the beam. This approach has been shown to be sufficient to measure and stabilize the accelerating voltage and phase by suppressing slow drifts in the RF system.

SwissFEL is a free electron laser facility, and its linear accelerator was built with TW structures operating at 100 Hz in pulsed mode [3]. For the upcoming upgrade, SwissFEL requires the acceleration of three bunches in the same RF pulse, enabling the simultaneous operation of three undulator lines at the full repetition rate. The accelerating voltages and phases of the three bunches are required to be measured and adjusted independently, leading to the request to develop a detailed TW-structure model. The model maps the measurable signals, such as the input and output waveforms of the TW structure and the beam current, to the accelerating voltage and phase for each bunch.

This paper summarizes the TW structure model and the derived control algorithms. Section II describes the TW structure model and the responses to RF drive power and beam loading. Section III discusses the control methods for TW structures based on feedforward, optimization, and feedback. In Section IV, the test results at SwissFEL are described to validate the TW-structure model. Finally, the outlook of the application of the TW structure model is discussed in Section V.

## II. TW STRUCTURE MODEL

The block diagram of a TW structure is shown in Fig. 1, where $L$ is the length of the structure, $L_c$ is the length of each cell (assuming the same for all cells), and $E$ is the electric field along the axis in the beam flight direction $z$.

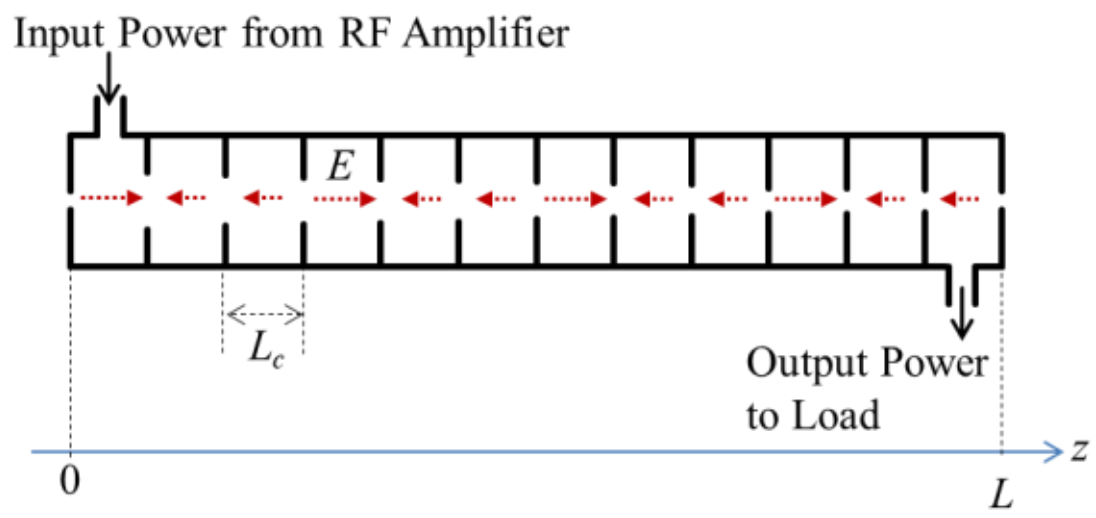

FIG 1. Block diagram of a TW structure.

The electromagnetic fields in the TW structure are excited by two inputs: the RF drive power and the beam. The RF signals at the structure input and load, as well as the beam current, can be measured in a practical accelerator. The accelerating voltage and phase experienced by the beam are not directly measurable and should be derived from these measurable signals. A TW structure is typically characterized by several quantities (see Table 1) in addition to the parameters shown in Fig 1.

TABLE 1. Characteristics of TW Structures [6].

| Parameters | Definition |
|---|---|
| Operating frequency | $\omega = 2\pi f$ |
| Shunt impedance per unit length | $r_s = \dfrac{E^2}{-dP_w/dz}$ |
| Quality factor | $Q = \dfrac{\omega W}{-dP_w/dz}$ |
| Normalized shunt impedance per unit length | $\rho := r/Q = \dfrac{E^2}{\omega W}$ |
| Group velocity | $v_g$ |

Here $E$ is the accelerating gradient, $P_w$ is the traveling wave power flowing in the structure, $-dP_w/dz$ is the resistive power dissipation per unit length in the structure walls, and $W$ is the stored electromagnetic energy per unit length in the structure. Note that $E$, $P_w$ and $W$ are functions of location $z$ and time $t$; and $r_s$, $Q$, $\rho$ and $v_g$ are all functions of $z$. The parameters in Table 1 are known design parameters of a TW structure. The RF drive power $P_d$ and drive voltage $V_d$ of the TW structure are functions of time and defined as

$$P_d = P_w(z=0),\ V_d = \sqrt{P_d} \tag{1}$$

where the impedance of the transmission line has been defined as 1 for simplicity. For the beam accelerated by the TW structure, we denote its bunch charge as $Q_b$, bunch spacing time as $T_b$, and then the average beam current can be defined as $I_0 = Q_b / T_b$. We assume that $T_b$ is constant, and $Q_b$ is a function of time (so does $I_0$).

One should be aware that $E$ and $V_d$ are the magnitudes of the RF quantities, and their time dependency does not include the oscillatory part at the RF frequency.

### A. Differential equation and general solution

Applying the definitions in Table 1, we obtain the differential equation for the accelerating gradient $E(z, t)$ based on the principle of energy conservation. Consider a small volume in the TW structure between $z$ and $z + \Delta z$. See Fig. 2.

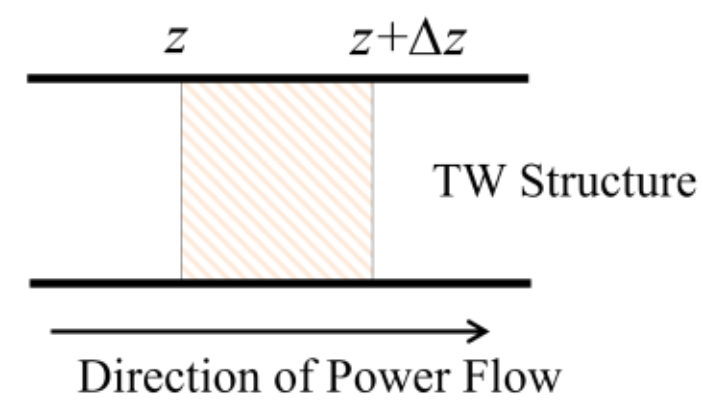

FIG 2. A small volume in the TW structure.

The change of stored energy in this volume from time $t$ to $t + \Delta t$ can be written as

$$\Delta U = \left[W(z,t+\Delta t) - W(z,t)\right]\Delta z \tag{2}$$

Here we assume that $\Delta z$ is very small compared to $L_c$, then, $W$ is approximately constant between $z$ and $z+\Delta z$. The change of stored energy originates from three sources: a). Energy flow of the traveling wave power $\Delta U_1 = \left[P_w(z,t) - P_w(z+\Delta z,t)\right]\Delta t$ ; b). Energy consumption on the resistive wall of the structure, calculated with the definition of $Q$ as $\Delta U_2 = -\dfrac{\omega W}{Q}\Delta z \Delta t$ ; c). Energy exchange with the beam, calculated as $\Delta U_3 = -EI\Delta z\Delta t$ , where $I(t)$ is the instantaneous beam current function (the independent variable $t$ is not shown in the formula for simplicity). According to the energy conservation theorem, we get $\Delta U = \Delta U_1 + \Delta U_2 + \Delta U_3$, leading to a differential equation if we take limits at $\Delta z$, $\Delta t \to 0$:

$$\frac{\partial W}{\partial t} = -\frac{\partial P_w}{\partial z} - \frac{\omega W}{Q} - EI \tag{3}$$

From the definition of group velocity, we get $P_w = v_g W$ , and its partial derivative for $z$ can be calculated as

$$\frac{\partial P_w}{\partial z} = v_g \frac{\partial W}{\partial z} + W\frac{dv_g}{dz} \tag{4}$$

From the definition of $\rho$ we get $W = E^2/(\omega\rho)$, so

$$\frac{\partial W}{\partial z} = \frac{1}{\omega\rho^2}\left(2E\rho\frac{\partial E}{\partial z} - E^2\frac{d\rho}{dz}\right), \qquad \frac{\partial W}{\partial t} = \frac{2E}{\omega\rho}\frac{\partial E}{\partial t} \tag{5}$$

By eliminating $W$ and $P_w$ with the above equations, we can get the differential equation of $E$ as

$$\frac{\partial E}{\partial t}+v_g\frac{\partial E}{\partial z}+\left(\frac{1}{2}\frac{dv_g}{dz}+\frac{\omega}{2Q}-\frac{v_g}{2\rho}\frac{d\rho}{dz}\right)E=-\frac{\omega\rho}{2}I \quad (6)$$

This is a partial differential equation with boundary conditions

$$E(0,t)=\sqrt{\frac{\omega\rho(0)P_d}{v_g(0)}}=\sqrt{\frac{\omega\rho(0)}{v_g(0)}}V_d, \quad (7)$$
$$E(z,0)=0$$

This partial differential equation has been solved in the paper by A. Lunin [7] and here we only give the result

$$E(z,t)=K(z)V_d(t-\tau(z))u(t-\tau(z))-G(z)\int_0^z I(t-\tau(z)+\tau(z'))u(t-\tau(z)+\tau(z'))\frac{\omega\rho(z')}{2v_g(z')G(z')}dz' \quad (8)$$

The items in the above solution are explained below:

a) The group delay of power transmission from the entry of the TW structure to the location $z$:

$$\tau(z)=\int_0^z\frac{dz'}{v_g(z')} \quad (9)$$

b) Attenuation functions:

$$G(z)=\sqrt{\frac{v_g(0)\rho(z)}{v_g(z)\rho(0)}}e^{-\int_0^z\frac{\omega}{2Q(z')v_g(z')}dz'},$$
$$K(z)=G(z)\sqrt{\frac{\omega\rho(0)}{v_g(0)}} \quad (10)$$

c) $u(t-t_0)$ is a unit step function at $t=t_0$.

The first term on the right-hand side of (8) is the RF-induced gradient, and the second term is the beam-induced gradient. To simplify the discussion of the relative phase between the overall RF field and the beam, we define the beam current phasor $\mathbf{i}=-I$ and the beam-induced gradient pointing in the negative direction (i.e., with a phase $\pi$), reflected by the minus sign before the second term of (8). It implies that we assume the beam current has a constant phase relative to the RF reference frequency, i.e., there is no bunch arrival time fluctuation, and the bunch repetition rate is critically locked to the RF frequency. It also implies that the cells of the TW structure are well tuned so that the beam-induced gradient has the same phase as the beam current phasor. Equation (8) shows the on-crest acceleration case, in which the phase of the drive signal $V_d$ is 0º. If we rotate the phase of the drive signal (e.g., introducing a phase of $\varphi_d$), we can write the overall RF gradient in the phasor form as

$$\mathbf{E}(z,t)=K(z)\mathbf{v}_d(t-\tau(z))u(t-\tau(z))-G(z)\int_0^z I(t-\tau(z)+\tau(z'))u(t-\tau(z)+\tau(z'))\frac{\omega\rho(z')}{2v_g(z')G(z')}dz' \quad (11)$$

where the input drive phasor is defined as

$$\mathbf{v}_d=V_d e^{j\varphi_d} \quad (12)$$

Then, the beam accelerating phase $\varphi_b$ is defined by the phase of the phasor $\mathbf{E}$, which is 0º if the beam is accelerated on crest. This relationship is depicted in Fig. 3. We use $\mathbf{E}_b$ to denote the beam-induced gradient with a phase of $\pi$, as defined previously. The RF drive-induced gradient is denoted as $\mathbf{E}_{rf}$ with the same phase as $\mathbf{v}_d$. Then the beam accelerating phase $\varphi_b$ is defined as the phase of $\mathbf{E}$, and therefore, the instantaneous electric field seen by the particle will be $|\mathbf{E}|\cos\varphi_b$.

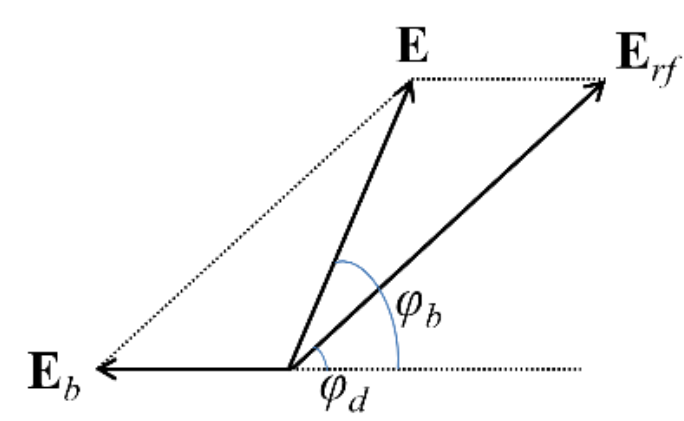

FIG 3. Vector diagram of the RF and beam-induced gradients (at the same location and time).

The RF-induced gradient (the first term of Eq. (11)) at location $z$ is produced by the RF drive power injected $\tau(z)$ earlier than the observation time and attenuated by the TW structure fraction between 0 and $z$. This is because the RF drive power is only coupled in from the start of the TW structure and fills the structure from start to end at the group velocity. When the RF power travels in the structure, it is attenuated due to the resistive power loss in the structure walls, and this process is modeled with $K(z)$.

Unlike the RF drive power, the beam passes through the TW structure and stimulates electromagnetic fields in all cells simultaneously. Note that we have neglected the beam flight time through the structure since it is much shorter than the filling time of the structure. This approximation is also supported by the purpose of the TW-structure module for studying and designing digital low-level RF (LLRF) controllers with a sampling rate of 250 MSPS. Any transient behaviors within one clock cycle (e.g., the beam flight time at the speed-of-light through a one-meter-long structure is about 3.3 ns) can be neglected and be viewed as a step change from the viewpoint of the LLRF system. The phase advance of the fields induced by the bunch in different cells matches well the TW-structure design in terms of cell lengths, i.e., if the TW-structure is well tuned, the fields induced by the same bunch along the structure will have the same phases when they arrive at a certain downstream location (e.g., at the load). Therefore, the beam-induced gradient at a location $z$ has two sources: a). Induced by the bunch that passes by $z$ at the observation time; b). The fields in upstream cells that are induced by earlier bunches and transmitted to $z$ at the observation time. This results in the integration in the formula of the beam-induced gradient (the second term of Eq. (11)).

### B. RF-induced accelerating voltage

The RF-induced gradient in the format of phasor can be written as

$$\mathbf{E}_{rf}\left(z,t\right)=K\left(z\right)\mathbf{v}_{d}\left(t-\tau(z)\right)u\left(t-\tau(z)\right) \quad (13)$$

By integrating it along the TW structure, we derive the accelerating voltage induced by the RF drive power as a function of $t$:

$$\mathbf{v}_{acc,rf}\left(t\right)=\int_{0}^{L}K\left(z\right)\mathbf{v}_{d}\left(t-\tau(z)\right)u\left(t-\tau(z)\right)dz \quad (14)$$

In practice, we measure the input and output of the TW structure with analog-to-digital converters (ADCs), which sample the waveforms at a constant sampling rate, resulting in equidistant samples. Therefore, integrating in time is more convenient in the software process. Therefore, we convert (14) from space-based integration to time-based integration. Equation (9) defines $\tau(z)$, which is the delay of RF power flowing from 0 to $z$ in the structure. We can define its inverse function as $z(\tau)$ to tell the location in the structure for a given delay. Then, if we replace the integration variable from $z$ to $\tau$, (14) can be rewritten as

$$\mathbf{v}_{acc,rf}\left(t\right)=\int_{0}^{T_{fill}}K\left(\tau\right)\mathbf{v}_{d}\left(t-\tau\right)u\left(t-\tau\right)\dot{z}\left(\tau\right)d\tau \quad (15)$$

where $T_{fill}$ is the filling time of the structure. The terms are defined as

$$T_{fill}=\tau\left(L\right),\ K\left(\tau\right)=K\left(z(\tau)\right),\ \dot{z}\left(\tau\right)=\frac{dz}{d\tau} \quad (16)$$

Assume we sample the waveforms at a rate of $f_s$, so the sampling time of each point is $T_s = 1/f_s$. Then, the integration in (15) can be calculated numerically from the sampling of the input signal as

$$\mathbf{v}_{acc,rf}\left(t\right)\approx\sum_{n=0}^{N_{fill}}\beta\left(n\right)\mathbf{v}_{d}\left(t-nT_{s}\right),$$

where (17)

$$\beta\left(n\right)=K\left(nT_{s}\right)\dot{z}\left(nT_{s}\right)T_{s},\ N_{fill}=\mathrm{int}\left(T_{fill}/T_{s}-1\right)$$

This is the formula for calculating the RF-induced accelerating voltage. We can see that the value of $\mathbf{v}_{acc,rf}$ at time $t$ is determined by the RF drive power from $t$-$T_{fill}$ to $t$. In practice, we are interested in the accelerating voltages at integer times of $T_s$; then, Eq. (17) can be directly used to calculate the values of $\mathbf{v}_{acc,rf}$ at $t = kT_s$ ($k$ = 0, 1, 2, …) using the samples of the RF input. The characteristics of the TW structure, such as $r_s$, $Q$, $\rho$ and $v_g$ for each cell, should be given by the structure designer as inputs for the calculation.

We use the SwissFEL C-band structure (parameters given in Fig. 4) as an example to study the TW structure model. The block diagram of a C-band RF station is depicted in Fig. 5. The 5712 MHz RF reference signal is delivered by the synchronization system, which is modulated in amplitude and phase by the low-level RF (LLRF) controller via an RF actuator consisting of two DACs and an I/Q modulator. The modulated signal is amplified by a pre-amplifier and then a klystron. The peak power of the klystron output is further increased by a barrel open cavity (BOC) to produce the RF drive power for the TW structures. We calculated the accelerating voltage and phase in the TW structure with Eq. (17). In the calculation, the input coupling factor and loaded quality factor of the BOC are set to 9.4 and 20165, respectively. The DAC pulse shape is the same as that applied to the C-band stations of SwissFEL. The input power to the BOC is assumed to be 33 MW, and the resulting maximum accelerating voltage in each structure (4 structures in a station) is about 63.4 MV. The waveforms of the simulation results are shown in Fig. 6.

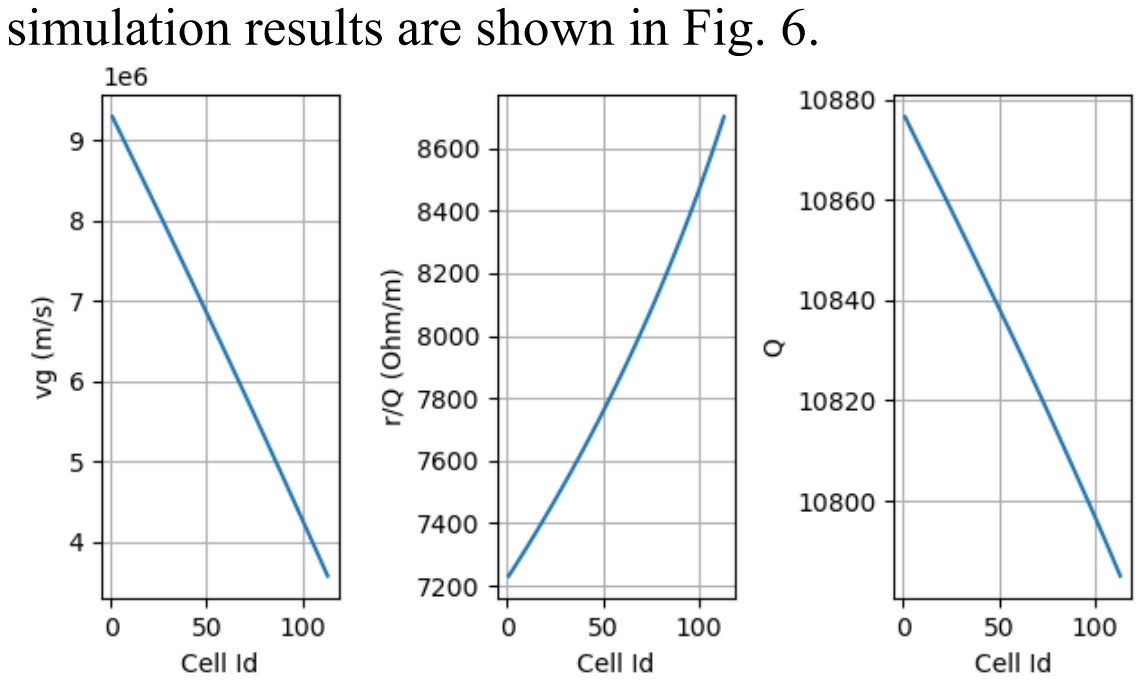

FIG 4. Parameters of the SwissFEL C-band TW structure in each cell.

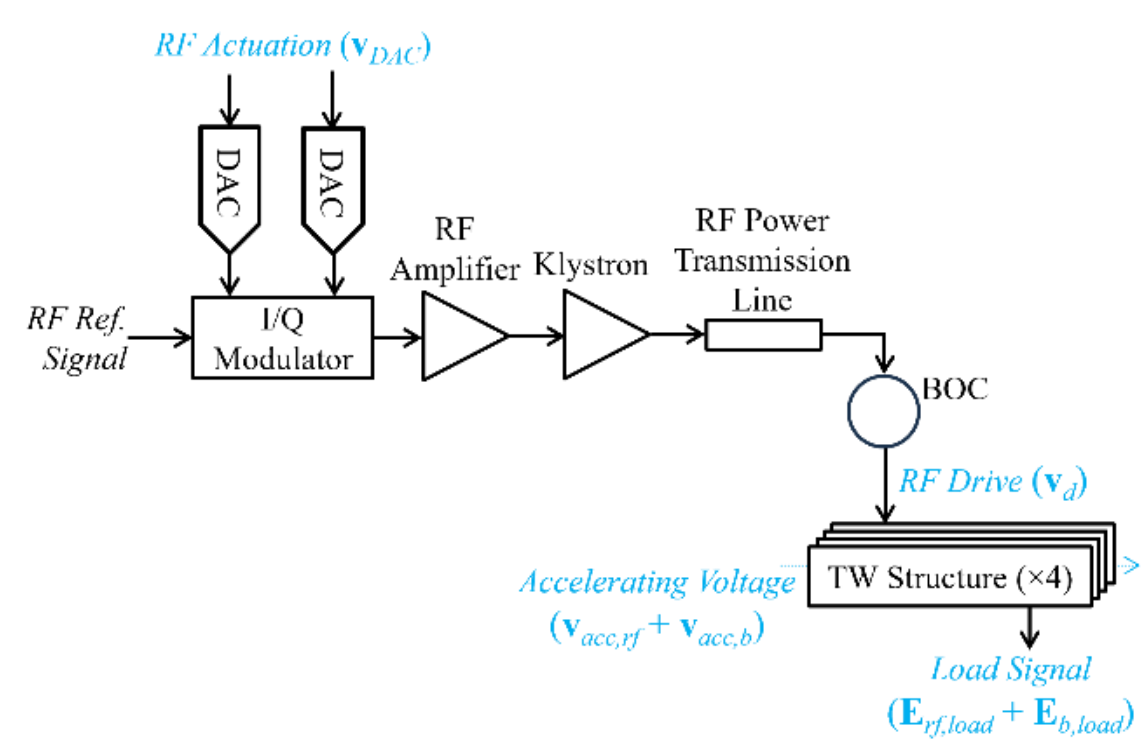

FIG 5. Block diagram of SwissFEL C-band RF station.

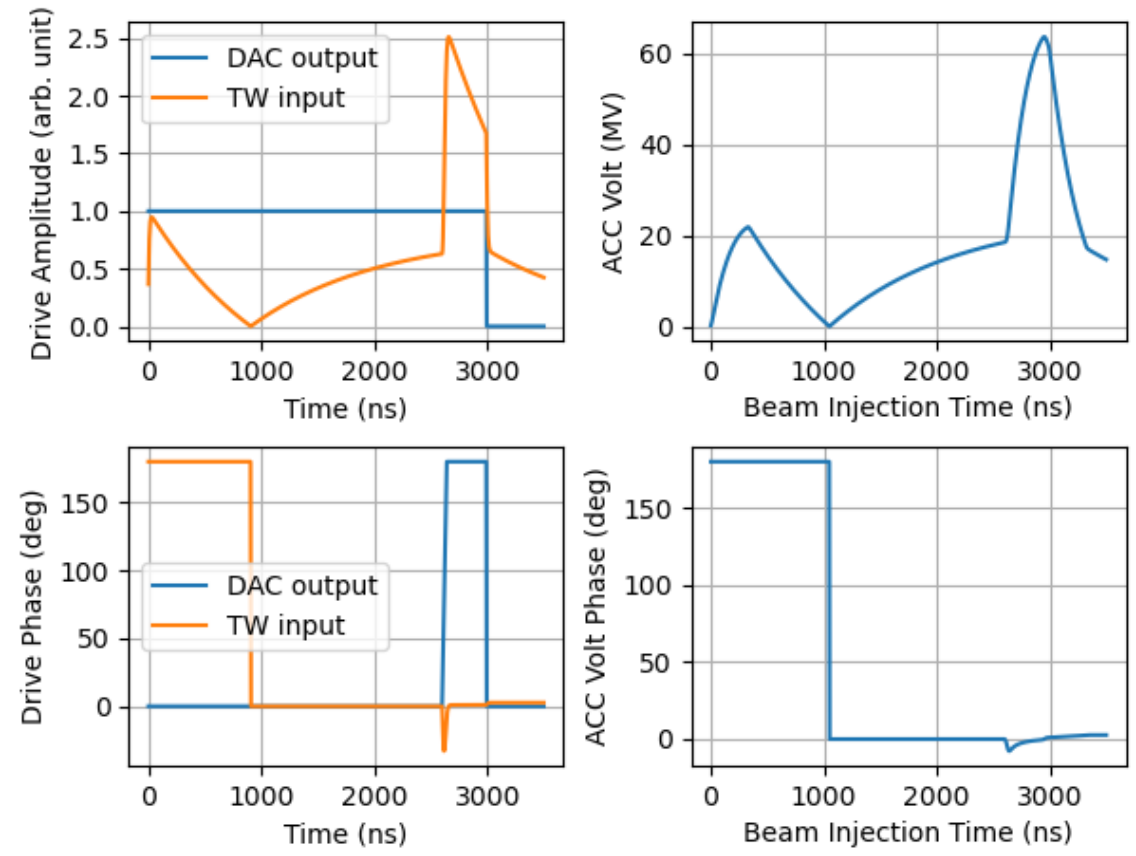

FIG 6. Accelerating voltage simulation for the typical operation of the SwissFEL C-band TW structure. The "Beam Injection Time" means if the beam is injected at this time, what accelerating voltage and phase the beam will experience.

In addition to the structure input signal, a TW structure is often equipped with a pickup at its load to measure the gradient at the end of the structure. According to Eq. (13), the load signal is a representation of the following quality

$$\mathbf{E}_{rf,load}(t) = \mathbf{E}_{rf}(L,t) = K(L)\mathbf{v}_d(t - T_{fill})u(t - T_{fill}) \quad (18)$$

It implies that the RF-induced load signal is a delayed version of the input signal $\mathbf{v}_d$ with the magnitude scaled by $K(L)$, the value of $K$ at the structure end.

Apply the Laplace transform to both sides of (17), we get the transfer function of the TW structure as

$$\mathbf{G}_{acc,rf}(s) = \frac{\mathbf{v}_{acc,rf}(s)}{\mathbf{v}_d(s)} = \sum_{n=0}^{N_{fill}} \beta(n) e^{-snT_s} \quad (19)$$

Its frequency response can be calculated by replacing $s$ with $j\omega$ with $\omega$ the angular frequency. As an example, Fig. 7 is the frequency response of the C-band TW structure of SwissFEL. The -3 dB bandwidth is estimated to be about 1.37 MHz. From the definition of $V_d$ as in (1), we can estimate that for an input power of 35 MW (square pulse), the achievable accelerating voltage is about 63 MV.

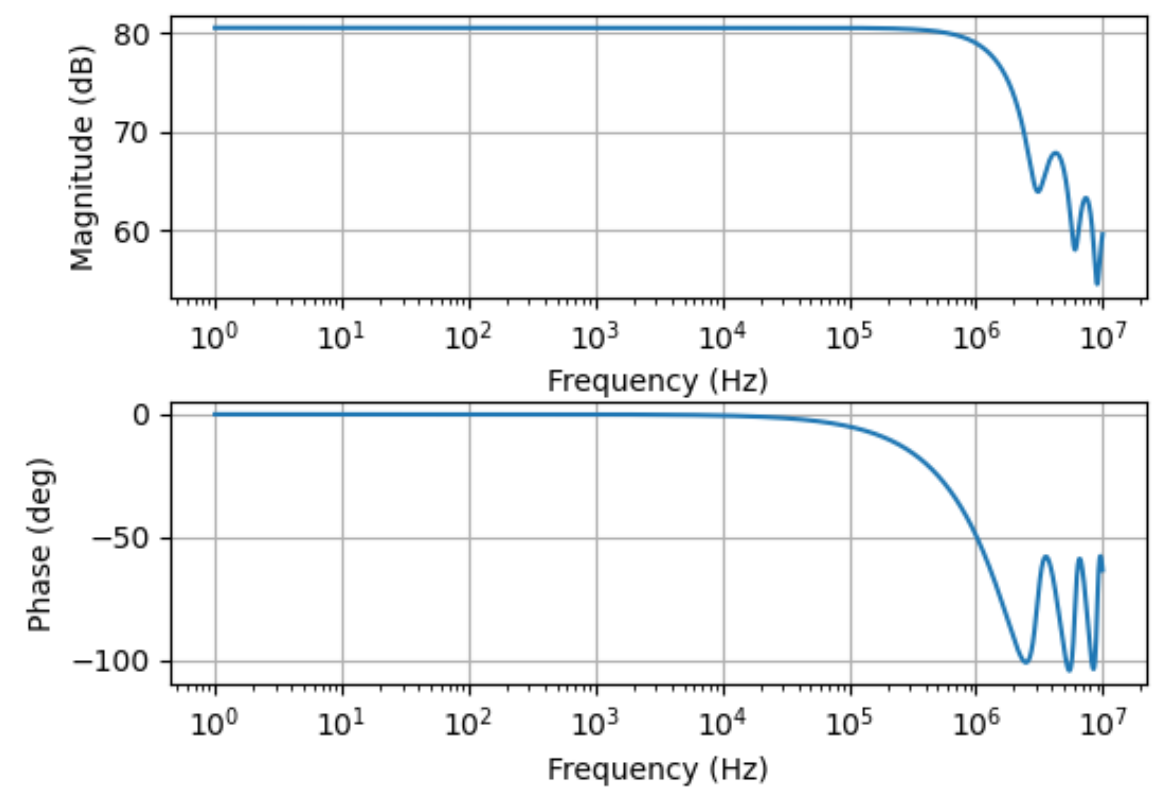

FIG 7. Frequency response of SwissFEL C-band TW structure (RF drive to accelerating voltage).

### C. Beam-induced accelerating voltage

The beam-induced gradient in the format of phasor can be written as

$$\mathbf{E}_b(z,t) = -G(z)\int_0^z I(t-\tau(z)+\tau(z'))u(t-\tau(z)+\tau(z'))\frac{\omega\rho(z')}{2v_g(z')G(z')}dz' \quad (20)$$

As mentioned before, its phase is defined as $\pi$, resulting in the minus sign in the formula. The accelerating voltage induced by the beam can be calculated by integrating (20) along the TW structure. With a similar treatment as in (14) and (15), the integration can be converted to be time-based as

$$\mathbf{v}_{acc,b}(t) = -\int_0^{T_{fill}}\left(G(\tau)\int_0^\tau I(t-\tau+\tau')u(t-\tau+\tau')\frac{\omega\rho(\tau')}{2v_g(\tau')G(\tau')}\dot{z}(\tau')d\tau'\right)\dot{z}(\tau)d\tau \quad (21)$$

The $T_{fill}$ and $\dot{z}(\tau)$ have been defined in (16) and the other terms are defined as

$$G(\tau) = G(z(\tau)),\ \rho(\tau) = \rho(z(\tau)),\ v_g(\tau) = v_g(z(\tau)) \quad (22)$$

Finally, the integration in (21) can be calculated numerically as

$$\mathbf{v}_{acc,b}(t) \approx -\sum_{n=0}^{N_{fill}}\left(\eta(n)\sum_{i=0}^{n}\xi(i)I(t-nT_s+iT_s)\right)$$

where

$$\eta(n) = G(nT_s)\dot{z}(nT_s)T_s,\ \xi(i) = \frac{\omega\rho(iT_s)\dot{z}(iT_s)T_s}{2v_g(iT_s)G(iT_s)} \quad (23)$$

Note that $N_{fill}$ has been defined in (17).

As an example, we show the typical accelerating voltage induced by the beam in the SwissFEL C-band TW structure. Assume we input an RF pulse with a flattop power of 10 MW and a phase of 30º. The RF pulse ramps up linearly as Fig. 8. After the RF pulse reaches the flattop and stays for the filling time of the TW structure (~ 320 ns), we inject a beam of 100 mA (average beam current). The resulting accelerating voltages by the RF input, the beam, and their vector sum are depicted in the right plots of Fig. 8. The phase of the beam-induced accelerating voltage is $\pi$, as we have defined before. The beam loading is clearly visible in both the amplitude and phase of the accelerating voltage: the total accelerating voltage is reduced by the beam loading, and the beam accelerating phase is shifted as well. It implies that if the beam loading (i.e., beam-induced gradient) is not negligible, it should be compensated for to achieve the desired overall accelerating voltage and beam accelerating phase.

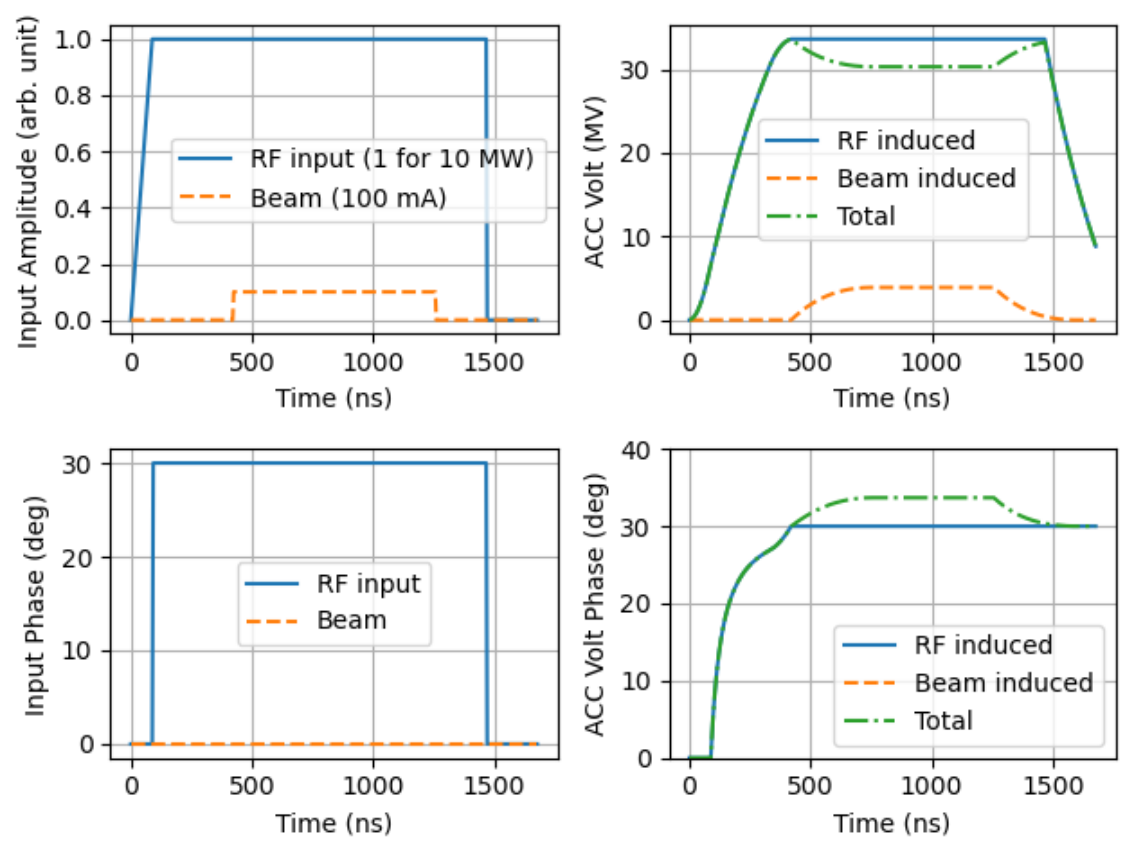

FIG 8. Example of beam-induced accelerating voltage in the C-band TW structure of SwissFEL.

In a TW structure, we can measure the beam-induced signal at the load of the structure, which corresponds to the beam-induced gradient at $z = L$ (or $\tau = T_{fill}$): $\mathbf{E}_{b,load}\left(t\right)=\mathbf{E}_{b}\left(L,t\right)$. Then, the beam-induced signal measured at the load can be presented as follows in terms of discrete integration:

$$\mathbf{E}_{b,load}\left(t\right)\approx -G\left(L\right)\sum_{n=0}^{N_{fill}}\xi\left(n\right)I\left(t-T_{fill}+nT_{s}\right) \quad (24)$$

The simulation of the beam-induced load signal of the SwissFEL C-band TW structure is shown in Fig. 9. The beam-induced signal at the load appears immediately when the beam is switched on. During the filling time of the structure, the signal ramps up, and then keeps constant until the beam is switched off. After the beam is switched off, the beam-induced load signal ramps down, corresponding to transmitting the beam-induced energy in the whole structure out of it.

In the above discussions, we have assumed that the instantaneous beam current is a square pulse with a constant flattop described by the function $I\left(t\right)=P\left(t,t_{1},t_{2}\right)I_{0}$, where $P(t, t_1, t_2)$ returns 1 in the time window between $t_1$ and $t_2$ and returns 0 in all other times. This is a typical approximation of the instantaneous beam current with a finite-length bunch train using the average beam current instead of the current of each bunch, resulting in an average response with the single-bunch-induced transients neglected.

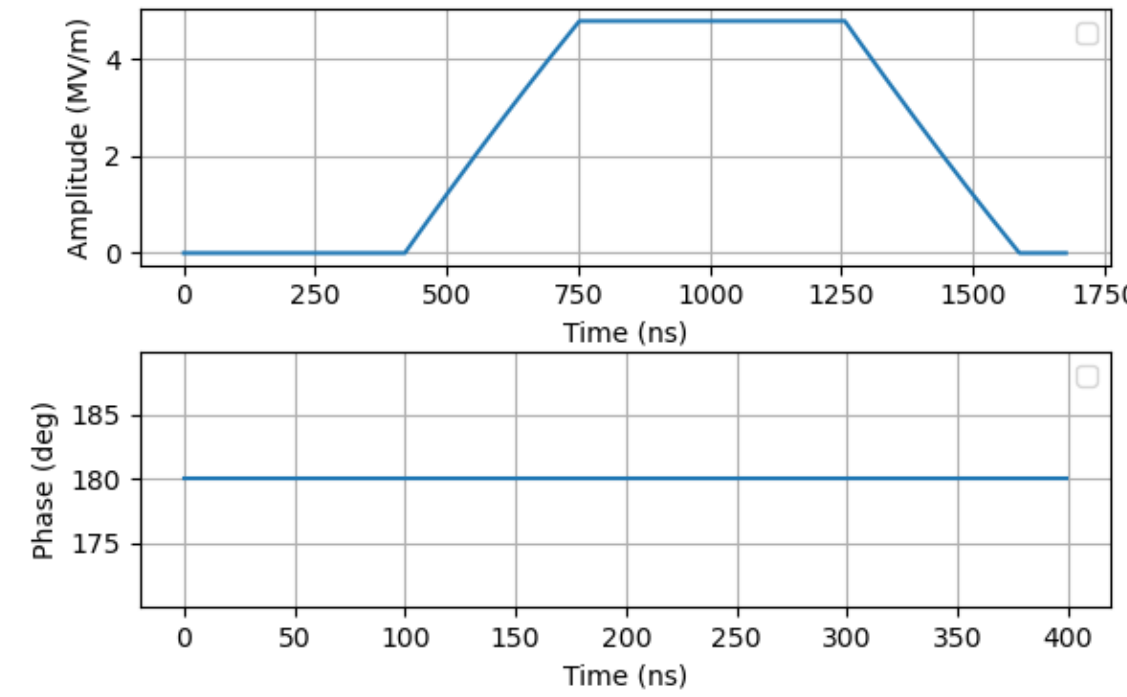

FIG 9. Beam-induced gradient at the load of the C-band TW structure of SwissFEL. The beam parameters are the same as in Fig. 8.

Apply the Laplace transform to both sides of (23), we get the transfer function of the TW structure for the beam current as

$$\mathbf{G}_{acc,b}\left(s\right)=\frac{\mathbf{v}_{acc,b}\left(s\right)}{I\left(s\right)}=-\sum_{n=0}^{N_{fill}}\left(\eta\left(n\right)\sum_{i=0}^{n}\xi\left(i\right)e^{-s(n-i)T_{s}}\right) \quad (25)$$

Here we use $I(s)$ to denote the Laplace transform of $I(t)$, which is the magnitude of the beam current phasor. The frequency response of $\mathbf{G}_{acc,b}(s)$ for the SwissFEL C-band TW structure is depicted in Fig. 10. The -3 dB bandwidth is estimated to be about 1.75 MHz, which is slightly larger than the bandwidth of the transfer function (19) with the RF drive as input.

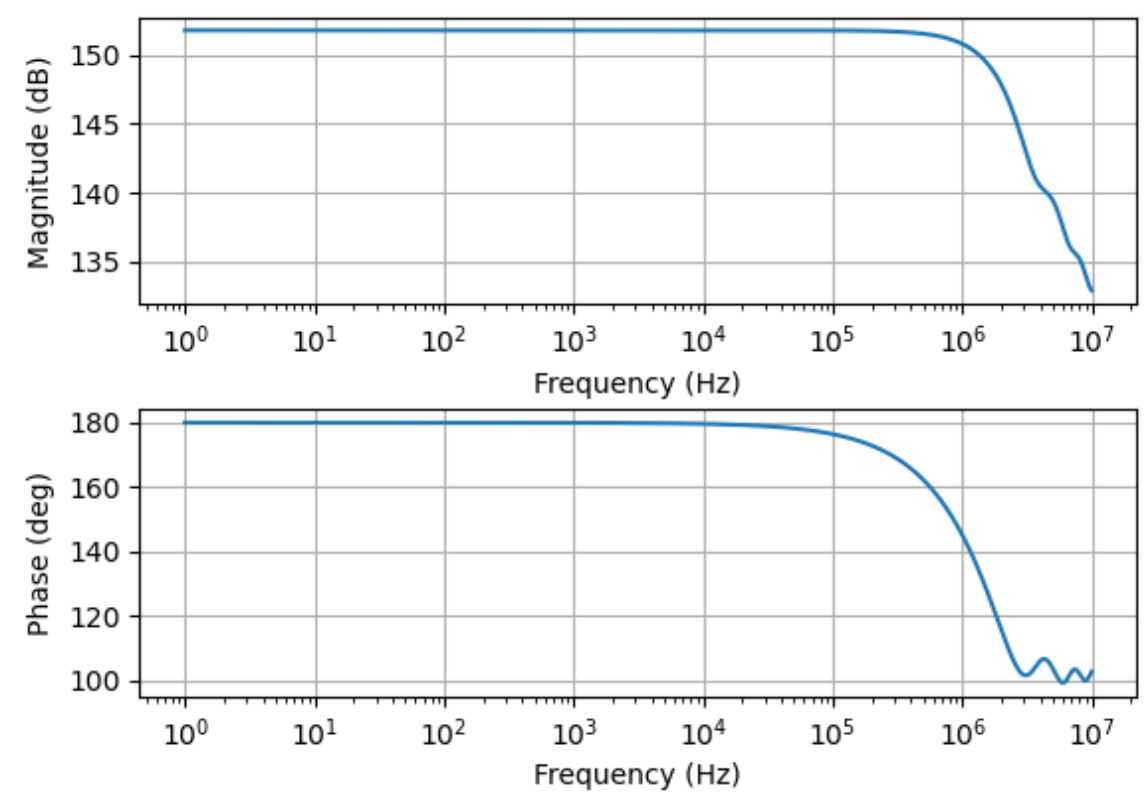

FIG 10. Frequency response of SwissFEL C-band TW structure (beam to accelerating voltage).

**D. Single-bunch-induced accelerating voltage**

In many accelerators based on TW structures, the beam loading is not that large, and only several bunches are present in the same RF pulse. In this case, it is useful to study the single-bunch-induced beam loading. For a single bunch injected at time $t = 0$, the instantaneous current can be written as

$$I\left(t\right)=Q_{b}\delta\left(t\right) \quad (26)$$

Put it into (21), we get the single-bunch-induced accelerating voltage as

$$\mathbf{v}_{acc,sb}(t) = -Q_b \int_0^{T_{fill}} G(\tau)\dot{z}(\tau)\frac{\omega\rho(\tau-t)\dot{z}(\tau-t)}{2v_g(\tau-t)G(\tau-t)}d\tau \tag{27}$$

The subscript "sb" stands for "single bunch". We calculate the integration with sum and only take the value of $\mathbf{v}_{acc,sb}$ at multiple of $kT_s$ ($k$ = 0, 1, 2, …), resulting in

$$\mathbf{v}_{acc,sb}(kT_s) \approx -\frac{Q_b}{T_s}\sum_{n=0}^{N_{fill}}\eta(n)\xi(n-k) \tag{28}$$

Note that we assign $\xi(n-k) = 0$ if $n < k$. In (28), the terms of $\eta$, $\xi$, and $N_{fill}$ have the same meanings as that in (23). For the single-bunch-induced load signal, we can calculate it with the method for deriving (24) by substituting the beam current with (26), leading to

$$\mathbf{E}_{sb,load}(kT_s) = -\frac{Q_b}{T_s}G(L)\xi(N_{fill}-k) \tag{29}$$

We take its values at $kT_s$, which are not zero only when $0 \le k \le N_{fill}$. As an example, we plot the single-bunch-induced accelerating voltage and load signal for the SwissFEL C-band TW structure as in Fig. 11. The bunch charge is assumed to be $Q_b$ = 4.2 nC injected at $t$ = 0.

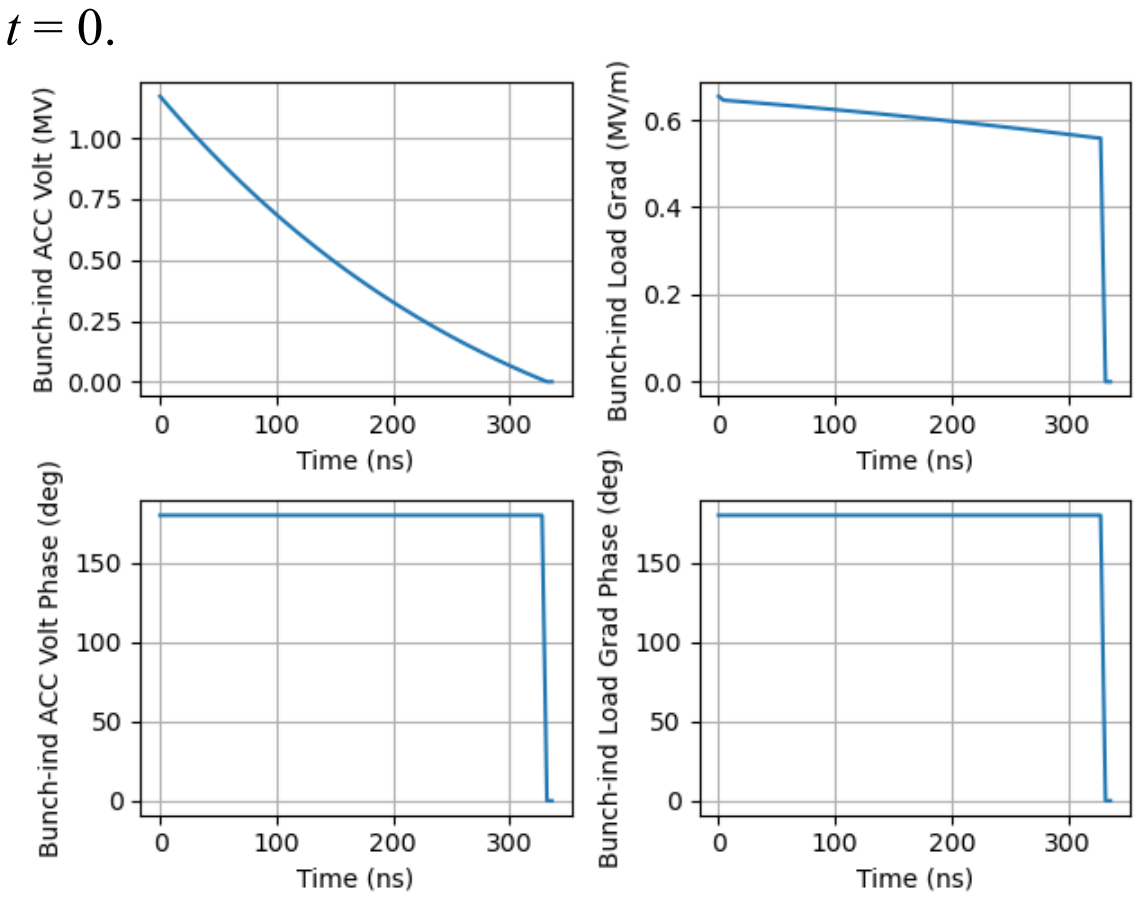

FIG 11. Single-bunch-induced accelerating voltage and load signal in SwissFEL C-band TW structure.

With the single-bunch-induced signals, the accelerating voltage and load signal induced by a bunch train can be calculated by summing up the contributions from all bunches with proper delays. We assume a bunch train with $Q_b$ = 4.2 nC and $T_b$ = 42 ns, corresponding to an average beam current $I_0$ = 100 mA, as in Fig. 8. Then, we sum up the single-bunch-induced signals from all bunches and derive the total beam-induced signals. The results are depicted in Fig. 12, compared with the beam-induced signals calculated with the average beam current. They agree well on average except for the transients of single bunches, which is as expected when approximating the bunch-train instantaneous beam current with the average current in the last subsection. This method can be used to calculate the beam-induced accelerating voltage for a beam with only a few bunches, or for bunches with different charges or phases.

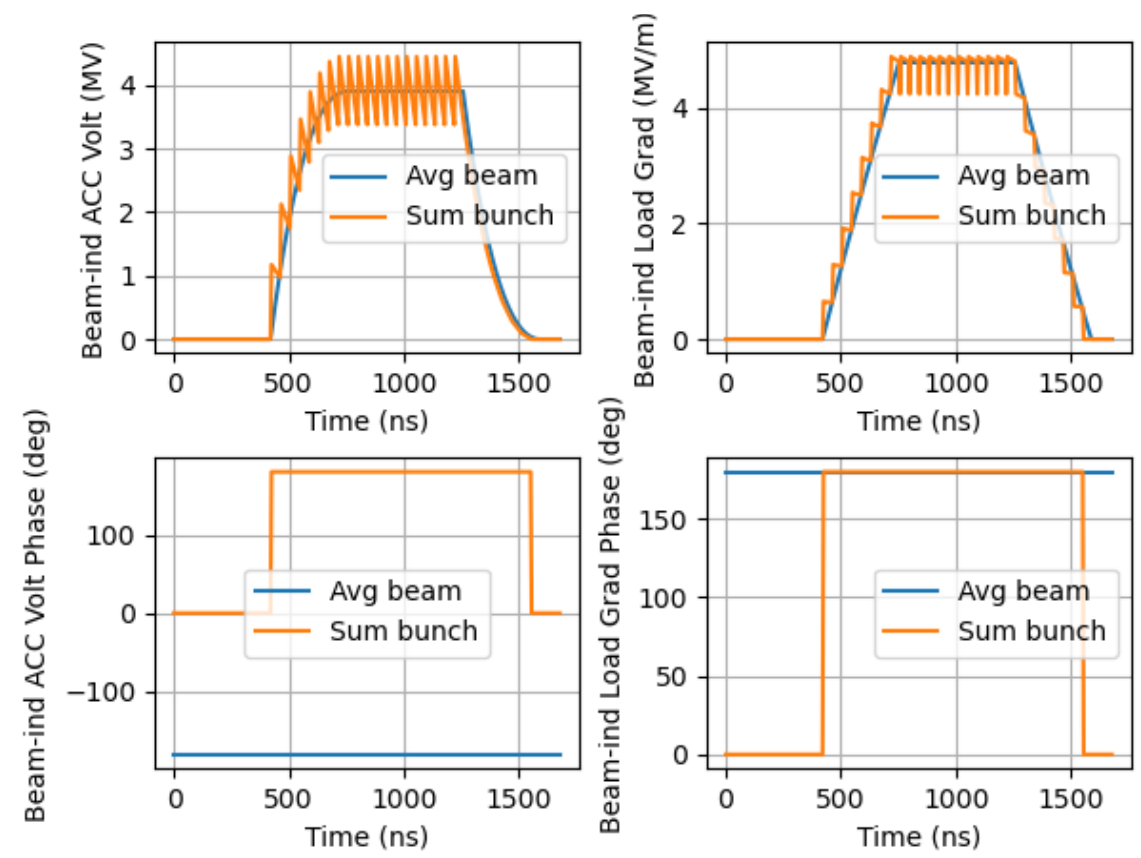

FIG 12. Construction of beam-induced signals by summing up the single-bunch-induced signals.

Modern high-resolution RF detectors of the LLRF system can measure the single-bunch-induced RF signal at the TW structure's load. The information derived from the measurement has multiple use cases together with the model developed here, including

1) *Determine the bunch injection time and RF-beam interaction region.*

The single-bunch-induced signal appears immediately in the load signal when the bunch passes through the TW structure. Here we ignore the bunch flight time through the structure since it is much smaller than the filling time of the structure. Therefore, the starting time of the single-bunch-induced signal indicates when the bunch passes through (i.e., the bunch injection time). This information is very useful to determine which part of the RF pulse should be used to calculate the accelerating voltage and phase experienced by the beam.

2) *Examine the phase balance of multiple structures.*

For RF stations with multiple TW structures driven by a single klystron, the RF field phases in all structures should be balanced (i.e., the same beam accelerating phase should be guaranteed for particles as they exit the upstream structure and enter the downstream one) to accelerate the particles most efficiently. The phase balance is typically optimized when installing the structures by carefully tuning the phase lengths of the waveguide network feeding the structures. In some cases, the phase balance between structures may have large errors due to the imperfect installation or other factors. We can use the single-bunch-induced beam loading measurement to roughly check the phase balance between different TW structures. To do that, we compare the phase of the single-bunch-induced load signal with the phase of the RF-induced load signal. The relative phase between them should be identical for all TW structures of the same RF station if they are well balanced.

*3) Examine the detuning of a TW structure.*

A TW structure may be detuned due to temperature errors. If it is detuned, there will be a slope in the absolute phase of the single-bunch-induced load signal. As we know, the starting point of the single-bunch-induced pulse is the gradient in the last cell of the structure, and the later points are the bunch-induced fields in upstream cells transmitted to the load. The fields in an upstream cell experience a phase shift when transmitting to the load, which is determined by the detuning of the structure and the location of the cell. The more upstream the cell is, the larger the phase shift is, resulting in a slope in the phase waveform of the single-bunch-induced load signal.

*4) Calibrate the TW structure RF drive power and accelerating voltage.*

If the bunch charge is known accurately, the theoretical absolute gradient induced by a bunch at the end of the TW structure can be calculated with (29). LLRF measures the RF signal picked from the structure load as a representation of the gradient (including contributions from RF and beam) with an arbitrary scale factor and phase offset compared to the actual value. Therefore, if we can measure the single-bunch-induced signal and the RF-induced signal at the TW structure load, we can use the theoretical value of the single-bunch-induced gradient to calibrate the measurements to physical units, e.g., the accelerating voltage in MV and the RF drive power in MW.

These applications based on the measurement of the single-bunch-induced load signal are tested at SwissFEL and the results are described in Section IV.

## III. TW STRUCTURE CONTROL METHODS

The control methods for RF stations equipped with TW structures can be studied with the model developed in the previous section. The following requirements need to be addressed:

1) Compensate for the beam loading by adding a feedforward term to the TW structure input.
2) Suppress random fluctuations in the RF system such as slow temperature drifts of the TW structure with feedback.
3) Manipulate intra-pulse accelerating voltages and phases for different bunches with feedforward.

The first two topics will be discussed in this section.

### A. Beam loading compensation

As shown by Fig. 8, if the TW structure accelerates a bunch train with a large average current, the beam-induced accelerating voltage (i.e., beam loading) will degrade the flatness of the total accelerating voltage seen by the bunch train. Such a beam loading is not directly measurable from the TW structure input and output signals, and therefore, cannot be suppressed by feedback. With the TW structure model, we can determine a feedforward correction to the TW structure input $\mathbf{v}_d$ to compensate for the beam loading. For a regular beam pulse (i.e., square pulse), analytical solutions can be found for the feedforward signal, whereas, for an arbitrary beam current (e.g., with gaps in the bunch train or for a bunch train with varying charges), numerical algorithms must be applied. We will discuss both cases.

#### *1. Square pulse beam current*

References [7-9] developed analytical solutions for the feedforward signal to compensate for the beam loading of a constant beam current switched on at a certain time. Here we follow a different method. We consider using the beam-induced load signal (24) as the feedforward signal. Let $\mathbf{v}_d(t) = -\mathbf{E}_{b,load}(t)$ and put into (17), we can calculate the resulting accelerating voltage as

$$\mathbf{v}_{acc,rf,FF}(t) = \sum_{n=0}^{N_{fill}} \beta(n) G(L) \sum_{i=0}^{N_{fill}} \xi(i) I(t - nT_s - T_{fill} + iT_s) \tag{30}$$

Here we have changed the index of the inner sum to $i$ to avoid confusion. Let's shift the waveform earlier by $T_{fill}$, that is, define $t' = t - T_{fill}$, and the above equation can be rewritten as

$$\mathbf{v}_{acc,rf,FF}(t' + T_{fill}) = K(L) \sum_{n=0}^{N_{fill}} \eta(n) \sum_{i=0}^{N_{fill}} \xi(i) I(t' - nT_s + iT_s) \tag{31}$$

Here we have used the relation between $K$ and $G$ in (10), and the definitions of $\beta$ and $\eta$ in (17) and (23), respectively. We assume the beam starts at $t = T_{fill}$, which is typical because the RF power needs to fill the TW structure before the beam is injected. Then, the relative timing between the beam current and the beam-induced load signal is explained in Fig. 13.

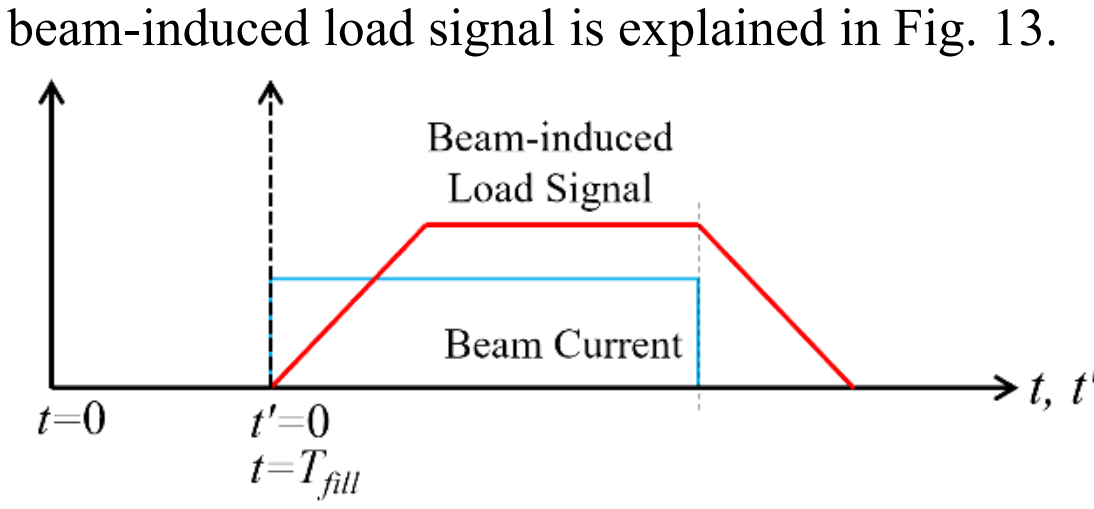

FIG 13. Relative timing between the beam current and the beam-induced load signal.

In the new time frame $t'$, the beam starts at $t' = 0$. Therefore, the index of the inner sum in (31) must be limited between 0 and $n$ for causality, leading to

$$\mathbf{v}_{acc,rf,FF}\left(t'+T_{fill}\right)= K\left(L\right)\sum_{n=0}^{N_{fill}}\eta\left(n\right)\sum_{i=0}^{n}\xi\left(i\right)I\left(t'-nT_s+iT_s\right) \quad (32)$$

It implies that if we use the load signal induced by a beam starting at $t' = 0$ as a feedforward correction signal ($\mathbf{v}_{d,corr}$) in the TW structure input, we can produce a correction of the accelerating voltage at $t' + T_{fill}$ as given by (32). Comparing (32) with the beam-induced accelerating voltage (23), we found that, except for a factor $K(L)$ and a different sign, the two formulas have the same format. Therefore, we can follow the procedure below to derive a feedforward correction signal to compensate for the beam loading of a square beam pulse:

1) Measure the beam-induced load signal, invert its sign, and divide it by $K(L)$.
2) Shift the waveform derived from step 1) to the left on the time axis by $T_{fill}$.
3) Extend the flattop of the waveform derived from step 2) to the end of the beam pulse and set the points after the beam to 0.
4) Add the waveform derived from step 3) to the TW structure input waveform $\mathbf{v}_d$.

The process above generates a flattop on the overall accelerating voltage covering the beam duration. We applied this algorithm to the case in Fig. 8, and Fig. 14 shows the results. The applied beam-induced load signal is shown in Fig. 9. The results indicate that the correction creates a flattop in the total accelerating voltage for both amplitude and phase. Note that the overall drive signal to the TW structure should be scaled and rotated to achieve the desired accelerating voltage and phase during the flattop (not shown here).

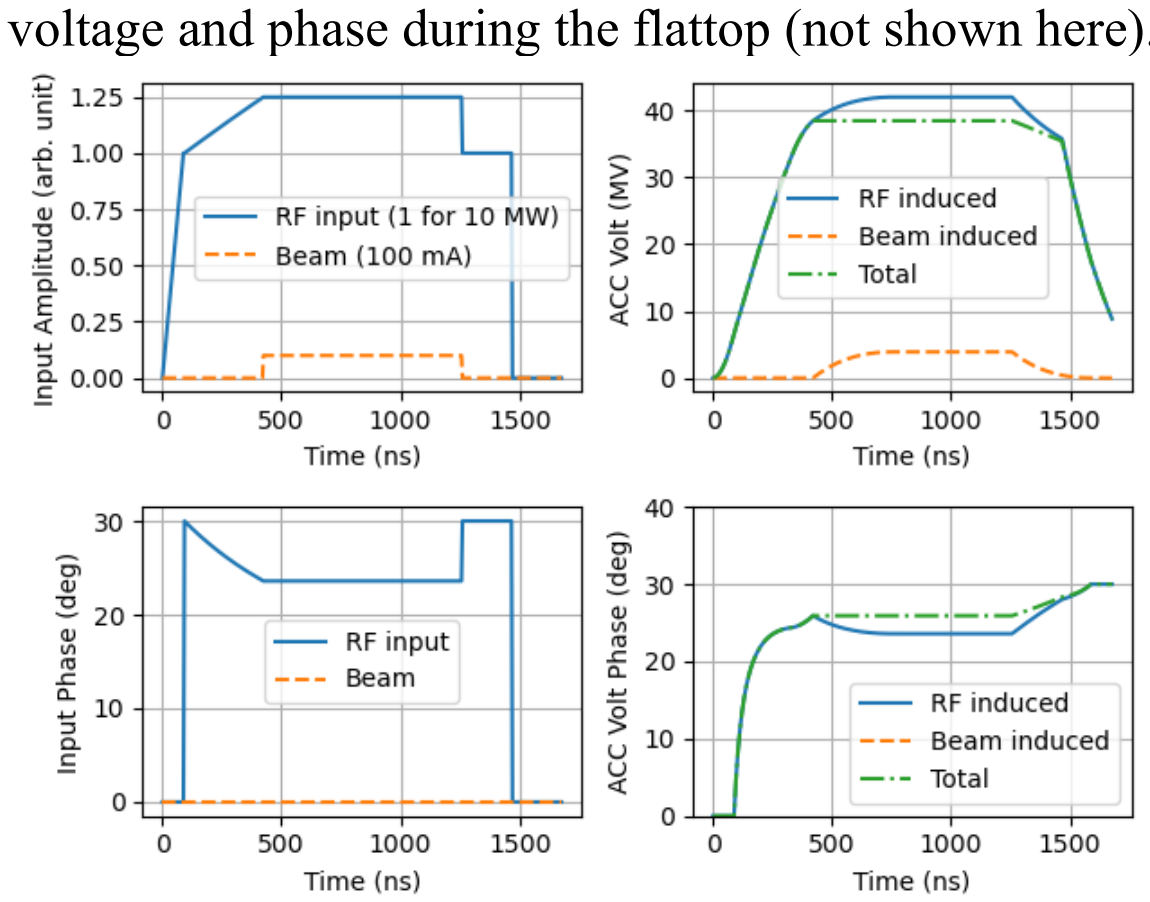

FIG 14. Beam loading compensation example for the case in Fig. 8.

***2. Arbitrary beam current***

The method introduced above has a limitation. It only works for a square-pulsed beam with a constant current at the flattop. The time interval between different beam pulses should be at least the TW structure filling time, leaving room to apply the feedforward signal.

If the beam current has arbitrary pulse shapes, it is very difficult to find accurate analytical solutions to compensate for the beam loading, and some attempts can be found in the reference [10]. Numerical methods should be used to determine the feedforward correction to the TW structure input. To do that, we need to convert the TW structure model (17) into a matrix format, i.e., construct a matrix $\mathbf{M}$ to transfer a vector of $\mathbf{v}_d$ to a vector of $\mathbf{v}_{acc,rf}$. If we only calculate the values of $\mathbf{v}_{acc,rf}$ at $kT_s$ with $k = k_0 + (0, 1, 2, \ldots)$, we can get the following formula

$$\begin{aligned}
\mathbf{v}_{acc,rf}\left(k_0\right) &= \beta\left(0\right)\mathbf{v}_d\left(k_0\right)+\beta\left(1\right)\mathbf{v}_d\left(k_0-1\right)+\cdots+\beta\left(N_{fill}\right)\mathbf{v}_d\left(k_0-N_{fill}\right)\\
\mathbf{v}_{acc,rf}\left(k_0+1\right) &= \beta\left(0\right)\mathbf{v}_d\left(k_0+1\right)+\beta\left(1\right)\mathbf{v}_d\left(k_0\right)+\cdots+\beta\left(N_{fill}\right)\mathbf{v}_d\left(k_0+1-N_{fill}\right)\\
&\vdots\\
\mathbf{v}_{acc,rf}\left(k_0+M-1\right) &= \beta\left(0\right)\mathbf{v}_d\left(k_0+M-1\right)+\beta\left(1\right)\mathbf{v}_d\left(k_0+M-2\right)+\cdots\\
&\quad+\beta\left(N_{fill}\right)\mathbf{v}_d\left(k_0+M-1-N_{fill}\right)
\end{aligned} \quad (33)$$

Here we dropped $T_s$ for simplicity by defining $\mathbf{v}_{acc,rf}\left(k\right):=\mathbf{v}_{acc,rf}\left(kT_s\right)$. The equation above can be rewritten in a matrix format as

$$\begin{aligned}
&\mathbf{V}_{acc,rf} = \mathbf{M}\mathbf{V}_d\text{, where}\\
&\mathbf{V}_{acc,rf} = \left[\mathbf{v}_{acc,rf}\left(k_0\right) \quad \mathbf{v}_{acc,rf}\left(k_0+1\right) \quad \cdots \quad \mathbf{v}_{acc,rf}\left(k_0+M-1\right)\right]^T \in \mathbb{C}^M,\\
&\mathbf{V}_d = \left[\mathbf{v}_d\left(k_0-N_{fill}\right) \quad \mathbf{v}_d\left(k_0+1-N_{fill}\right) \quad \cdots \quad \mathbf{v}_d\left(k_0+M-1\right)\right]^T \in \mathbb{C}^{M+N_{fill}},\\
&\mathbf{M} = \begin{bmatrix} \beta\left(N_{fill}\right) & \cdots & \beta(1) & \beta(0) & & & \\ & \beta\left(N_{fill}\right) & \cdots & \beta(1) & \beta(0) & & \\ & & \ddots & & & \ddots & \end{bmatrix} \in \mathbb{R}^{M\times\left(M+N_{fill}\right)}.
\end{aligned} \quad (34)$$

It implies that $M + N_{fill}$ points of $\mathbf{v}_d$ determine $M$ points of $\mathbf{v}_{acc,rf}$.

To compensate for the beam loading, we need to find the correction signal $\mathbf{v}_{d,corr}$ to cancel the beam-induced accelerating voltage $\mathbf{v}_{acc,b}$. This leads to the following optimization problem

$$\mathbf{V}_{d,corr} = \underset{\Delta\mathbf{V}_d}{\arg\min}\left\|\mathbf{V}_{acc,b}+\mathbf{M}\Delta\mathbf{V}_d\right\|_2 \quad (35)$$

Here $\mathbf{V}_{acc,b}$ is the vector of the beam-induced accelerating voltage with the same format as $\mathbf{V}_{acc,rf}$ in (34), and $\Delta\mathbf{V}_d$ is the incremental vector of the structure input. This optimization problem is convex and can be solved with the software tools of CVX in MATLAB [11] or CVXOPT in Python [12]. We applied this algorithm to the same problem as in the last subsection, except that the beam current has a gap (with half of the current) within the square pulse. The results are shown in Fig. 15. Compared to Fig. 14, the correction signal to the TW structure input is quite irregular.

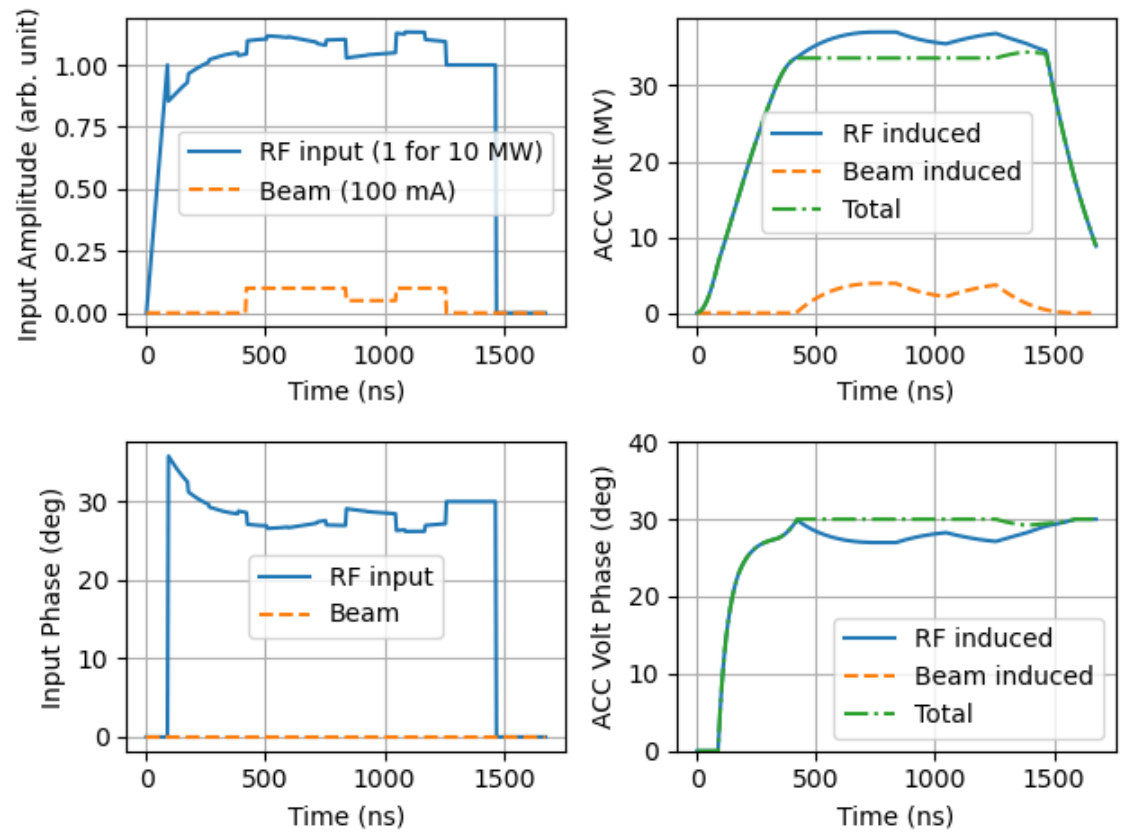

FIG 15. Compensation of arbitrary beam current.

### *3. Practical considerations*

The beam loading compensation (BLC) algorithms above are explained with simulations by directly correcting the TW structure input signal. To apply the BLC algorithms in a practical RF control system, we need to determine the correction to the RF controller output ($\mathbf{v}_{DAC}$) instead of the TW structure input ($\mathbf{v}_d$). For this purpose, we can write the transfer function between the LLRF controller output ($\mathbf{V}_{DAC}$, as a vector with the same format as $\mathbf{V}_d$) and the structure input $\mathbf{V}_d$ in the matrix format as

$$\mathbf{V}_d = \mathbf{G}\mathbf{V}_{DAC} \tag{36}$$

where $\mathbf{G}$ is determined by the gains, phase shifts, and dynamic responses of the RF driving chain components between the LLRF controller and the TW structure input, such as the up-conversion mixer, RF amplifier, BOC, etc. The response matrix $\mathbf{G}$ can be derived from the impulse responses of the RF driving chain components, as described in Subsect. 4.5.2 of the LLRF book [4]. Then, if we derive the correction to the TW structure input $\mathbf{V}_{d,corr}$, the correction to the RF controller output can be calculated by solving the optimization problem

$$\mathbf{V}_{DAC,corr} = \arg\min_{\Delta\mathbf{V}_{DAC}} \left\| \mathbf{V}_{d,corr} - \mathbf{G}\Delta\mathbf{V}_{DAC} \right\|_2 \tag{37}$$

For the "arbitrary beam current" case, the optimization can be combined with (35) to directly calculate the correction to the DAC output by solving

$$\mathbf{V}_{DAC,corr} = \arg\min_{\Delta\mathbf{V}_{DAC}} \left\| \mathbf{V}_{acc,b} + \mathbf{M}\mathbf{G}\Delta\mathbf{V}_{DAC} \right\|_2 \tag{38}$$

Another practical problem we might meet is the dispersion in the TW structure, which has been neglected in the model developed in this paper. The dispersion of the TW structure distorts the input RF pulse and causes additional amplitude or phase errors for different parts in the bunch train. A similar method, as described in this section, can also be used to correct the dispersion effects if we include the dispersion effects in the model, which has been discussed in reference [13]. In practice, the model developed in this paper should be sufficient if we apply it iteratively to control the bunch train according to the measurements of the beam parameters (e.g., energy) of each bunch.

### B. RF feedback control

As discussed in previous sections, a TW structure has the following features:

1) The accelerating voltage and phase experienced by the beam can be calculated with the structure input signal using the TW-structure model. We should take the correct part of the structure input pulse interacting with the beam for the calculation. The structure input signal is typically measured by LLRF with a pickup installed in the waveguide before the TW structure input coupler.
2) The load signal is also often measured. However, the load signal is complex in the presence of a beam. If there is no beam, the phase difference between the load signal and the TW structure input signal reflects the phase drift of the structure. We assume the phase drift of the TW structure is distributed linearly in the structure.

Based on the analysis, a feedback control scheme is proposed as in Fig. 16. Here, we assume that the beam loading has been well compensated for by the "RF Pulse Actuation" block implementing the feedforward algorithms discussed before.

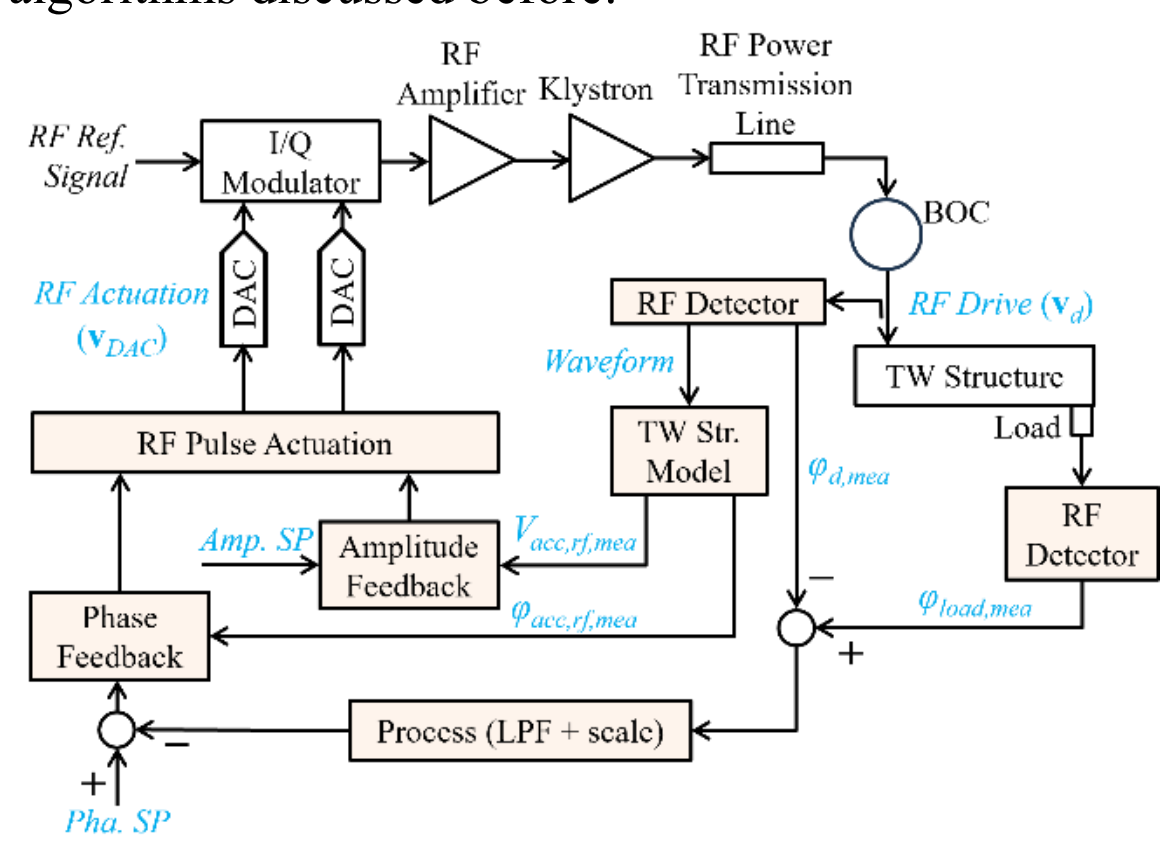

FIG 16. Proposed feedback control scheme for RF stations with TW structures.

The amplitude and phase feedback (e.g., pulse-to-pulse feedback) is based on the TW structure input signal $\mathbf{v}_d$, with the accelerating voltage and phase calculated with the TW structure model. To compensate for the drift in the TW structure, we measure the phase difference between the structure output and input, then the variation of this phase difference is processed to determine the rotation of the RF station's phase by modifying the setpoint of the phase feedback loop. Note that the phase of the load signal should be measured from where the beam-induced signal is not present. For example, assuming the feedback has well stabilized the structure input, and the structure has a phase drift of 1 degree (distributed linearly along the structure). The "Process (LPF + scale)" module detects this phase drift and subtracts 0.5° from the phase setpoint. If we leave the phase drift uncontrolled, the beam will see a phase error from 0° to 1° when it passes through the structure. With the phase setpoint correction, the phase feedback loop rotates the RF station's phase by –0.5°, causing the phase error experienced by the beam to change from –0.5° to 0.5° when passing through. This makes the overall beam accelerating phase unchanged except for a slight reduction in accelerating voltage, which is typically not critical and can be compensated for by beam-based feedback loops.

## IV. SWISSFEL TEST RESULTS

This section presents the test results at SwissFEL to validate the models and algorithms developed in this work.

### A. Application of the single-bunch-induced signal measured at the load

The resolution of the SwissFEL LLRF system is sufficiently high to measure the single-bunch-induced beam loading. The single-bunch-induced load signal can be derived by calculating the difference between the waveforms with and without beam. To increase the signal-to-noise ratio, we can collect many pulses in both cases and average them. Fig. 17 shows the detection of the beam loading signal induced by a 200-pC electron bunch in the TW structures of the first C-band RF station of Linac1 (S10CB01) at SwissFEL.

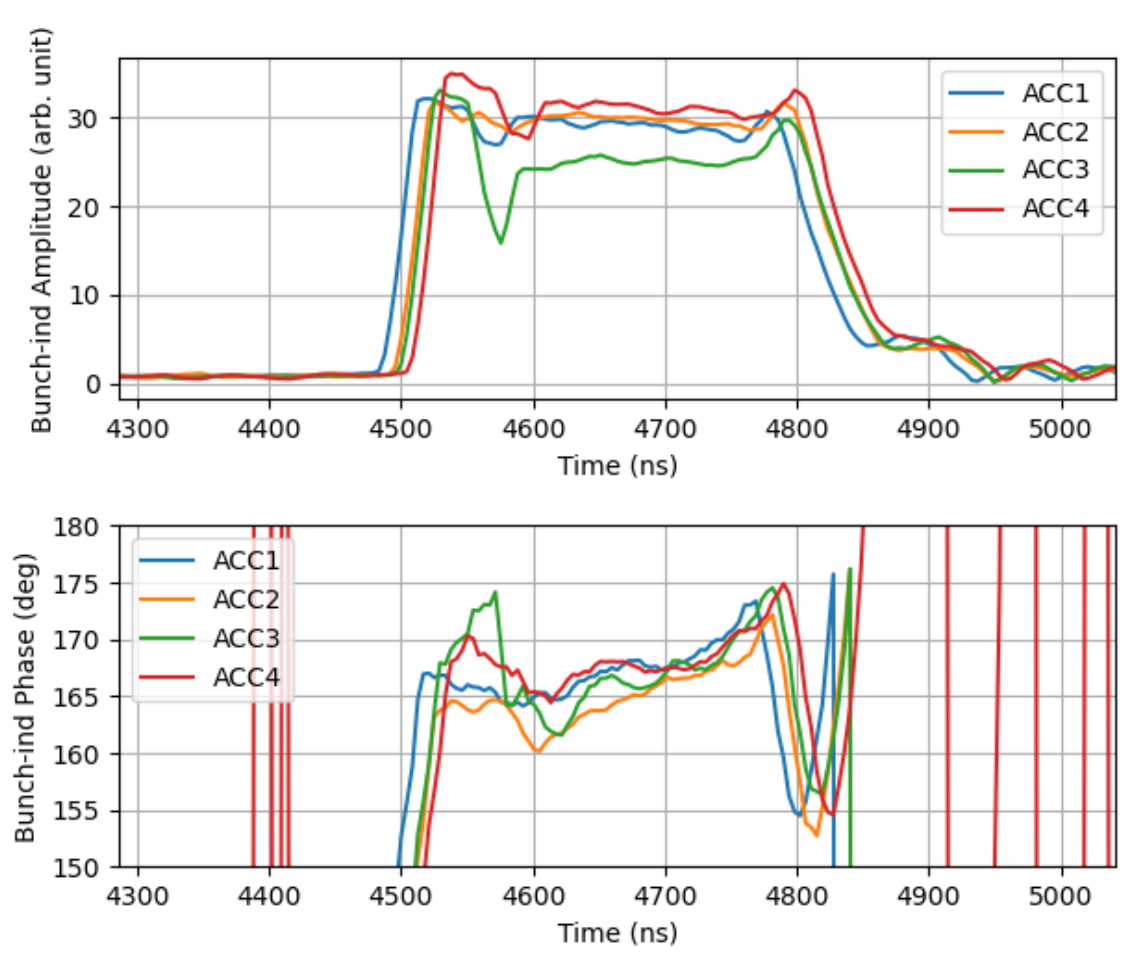

FIG 17. Single-bunch-induced load signals in the four TW structures of S10CB01.

As discussed in section II.D, the single-bunch-induced load signals can be used in various applications. They are described as follows.

1) *Determine the bunch injection time and RF-beam interaction region.*

The rising edges of the single-bunch-induced load signals indicate the beam injection times into different TW structures. If we compare the beam injection time with the waveform of the TW structure input signal (assuming the cables picking up the input and load signals of a structure have the same length), the region in the TW structure input pulse interacting with the beam can be identified, which starts from the beam injection time and has a length of the structure's filling time.

2) *Examine the phase balance of multiple structures.*

At SwissFEL, the phases of the 4 load signals of S10CB01 have been calibrated to the same reference plane. Therefore, we can directly compare the relative phases of the single-bunch-induced load signals to check the phase balance across the 4 structures. From the results in Fig. 17, the phases are close to each other, implying acceptable phase balance among the 4 structures. Note that this method can only detect significant phase imbalances, since the measurements of single-bunch-induced signals are often noisy due to small signal levels.

3) *Examine the detuning of a TW structure.*

As shown in the phase waveforms in Fig. 17, a total phase shift of about 10º is observed in the structures, implying that the cooling-water temperature can be further optimized to increase the acceleration efficiency (i.e., when the phase waveforms become flat, the acceleration efficiency is maximized).

4) *Calibrate the TW structure RF drive power and accelerating voltage.*

As discussed previously, we can estimate the RF drive power and accelerating voltage in physical units from the single-bunch-induced load signal if the bunch charge is known. The bunch charge for SwissFEL is 200 pC. The induced load signal in a C-band TW structure can be estimated to be around 30 kV/m by scaling the results in Fig. 11. Compared with the amplitude waveform in Fig. 17, we can calibrate the load signal measurement to a physical unit in MV/m. Then, with equations (18), (17), and (1), we can calculate the magnitude of $\mathbf{v}_d$, the magnitude of $\mathbf{v}_{acc,rf}$, and the TW structure input power in physical units. Figure 18 shows the estimate of the RF drive power for the first structure of S10CB01, and Fig. 19 is the accelerating voltage.

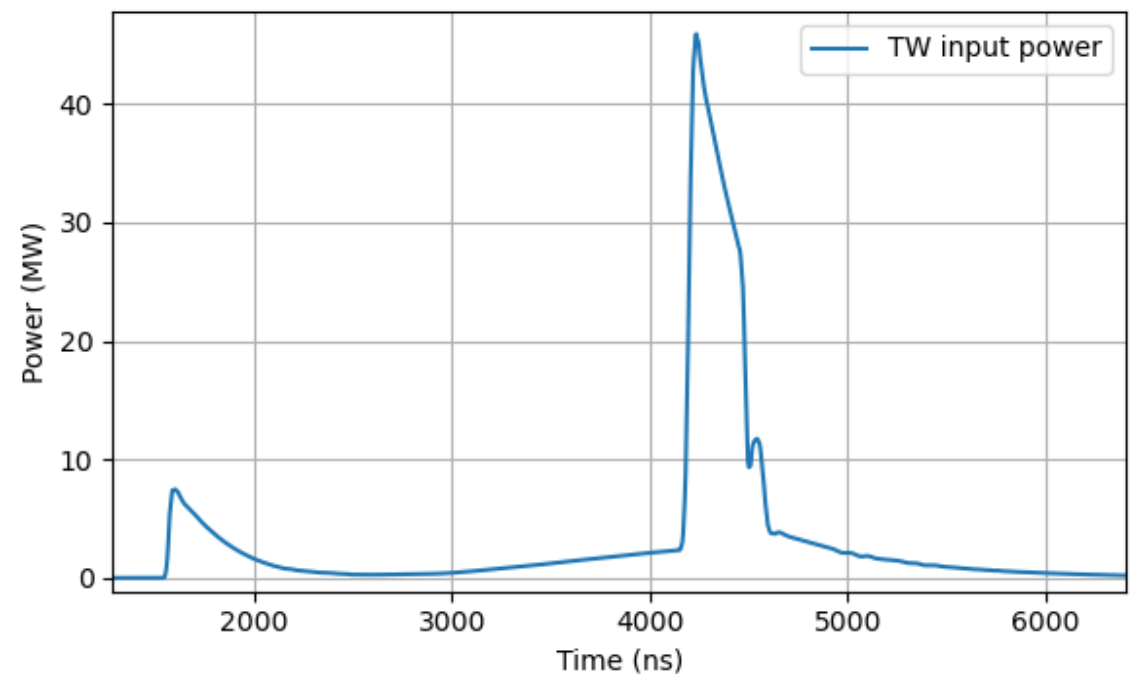

FIG 18. Estimate of the RF drive power for the first structure of S10CB01.

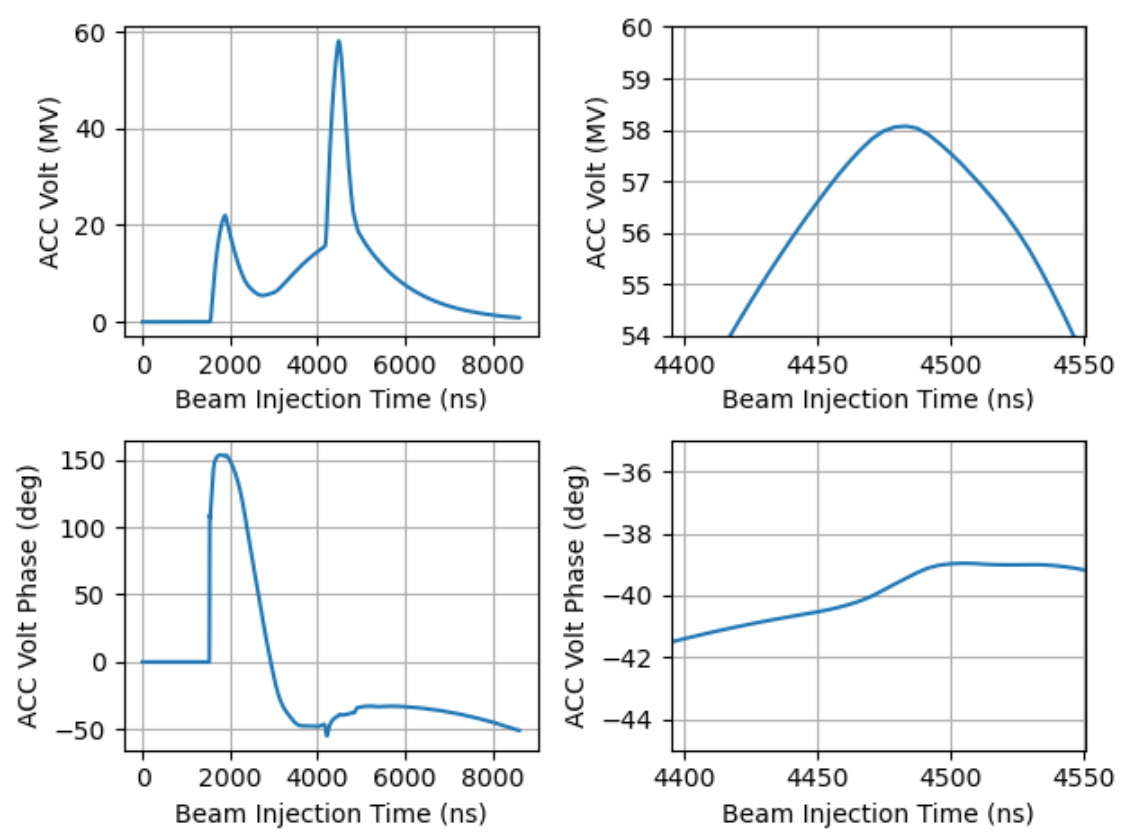

FIG 19. Estimate of the accelerating voltage curve for the first structure of S10CB01. The right plots show zoomed views of the left plots. The total accelerating voltage of S10CB01 with 4 TW structures is 230 MV according to the beam energy-based calibration, which is in good agreement with the estimate here.

The calibration here relies on several factors: (a). the measurement of the bunch charge is accurate; (b). the TW-structure model is accurate; (c). the measure of the load signal is accurate. The calibration of the accelerating voltage is more accurate with the beam energy changes when scanning the RF phase or applying/bypassing the RF station [5].

### B. TW structure model validation with beam

There are several ways to validate the TW structure model. The basic method is based on the beam energy measurement. We assume that the structure input power is known, which can be estimated from the power measurement of the structure input pickup signal, the path attenuation, and the coupling factor of the directional coupler. Then the accelerating voltage of the RF station can be calculated with the TW structure model. By comparing the estimated accelerating voltage with the beam energy gain in the case of on-crest acceleration, the validity of the TW structure model can be examined. This was demonstrated in Fig. 19.

Another method to validate the TW structure model is to evaluate beam stability under RF feedback control. Pulse-to-pulse RF feedback stabilizes the RF fields for stable beam acceleration. Therefore, the RF feedback effects on beam stability strongly depend on how accurate the RF field measurements can reflect the RF fields experienced by the beam. For SwissFEL, there are two factors influencing the accuracy of the accelerating voltage and phase measurements: (a). selecting the correct region in the TW structure input pulse that interacts with the beam; (b). applying the TW structure model to calculate the accelerating voltage and phase. If these factors are correctly addressed, the RF feedback should yield the optimal effect on beam stability.

This method was tested at S10CB01of SwissFEL. Pulse-to-pulse feedback was applied to the vector sum [14] of the inputs from the 4 TW structures. The RF-beam interaction regions within the 4 TW structure input signals were identified with the single-bunch-induced signals, as discussed in the previous subsection. The beam energy was measured at the downstream bunch compressor with a beam position monitor (BPM) installed in the dispersive region. To observe the beam stability affected by the RF feedback more clearly, we injected random amplitude and phase disturbances in S10CB01, resulting in beam energy jitter much larger than the natural jitter (e.g., caused by RF field jitter of other stations, beam arrival time jitter, orbit jitter, etc.) without artificial disturbances. The beam energy jitter was measured under different conditions:

1) RF feedback in open loop, without artificial disturbances (labeled as “OL no dist”).
2) RF feedback in open loop, with artificial disturbances (labeled as "OL dist”).
3) RF feedback in closed loop, RF measurement in correct time window (RF-beam interaction region), and with TW structure model applied (labeled as “FB CT TW”). The “RF-beam interaction region” is a time window on the input RF pulse. The RF power within this time window builds up the fields in all cells of the TW structure when the bunch passes through the structure. A bunch arriving earlier or later experiences RF fields built by the RF drive power within different time windows [5].
4) RF feedback in closed loop, RF measurement in correct time window, and with simple average (labeled as “FB CT Avg”). This is the current configuration of SwissFEL RF stations.
5) RF feedback in closed loop, RF measurement in a shifted time window (10 samples to left of the RF-beam interaction region), and with TW structure model applied (labeled as “FB ST TW”).

The resulting beam energy jitters for these 5 configurations are depicted in Fig. 20. As expected, if we choose the correct RF pulse region and apply the TW structure model, the beam can be stabilized to a level very close to the case without artificial disturbances. The result is a good indication of the validity of the TW structure model developed in this paper.

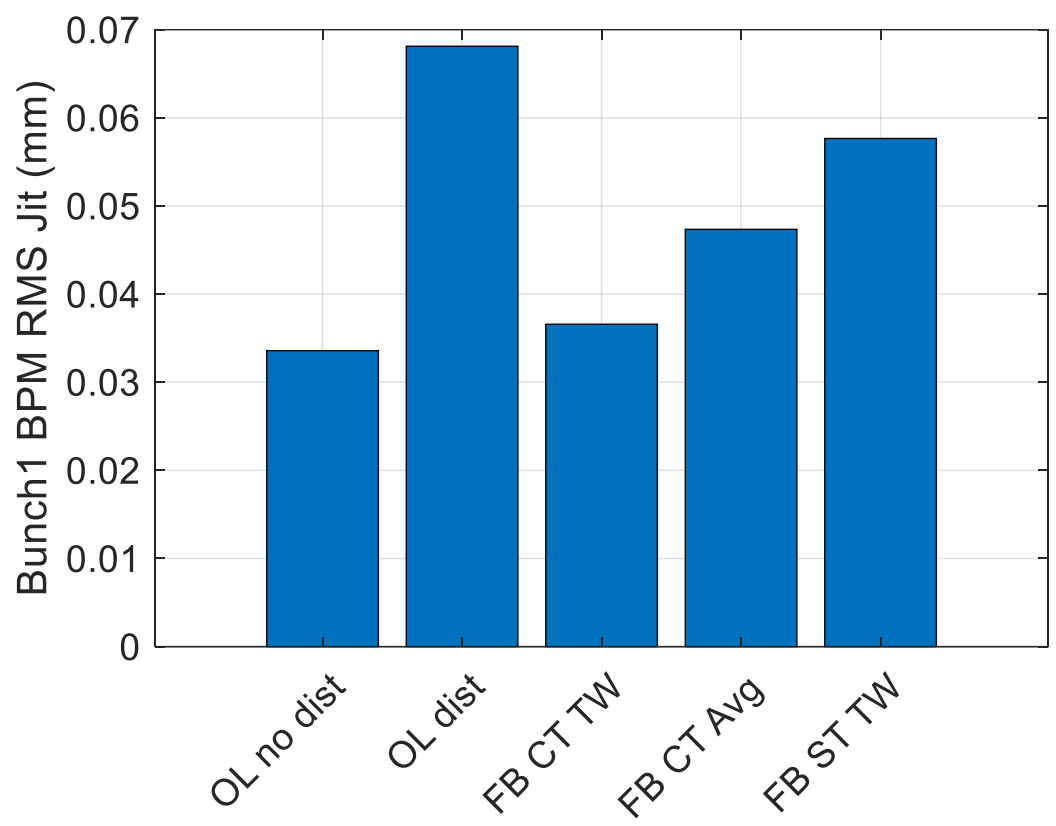

FIG 20. Beam energy stability with different RF feedback configurations.

## V. CONCLUSION AND OUTLOOK

This paper systematically discusses the modeling and control methods for TW structures. The differential equation of the structure and its general solution were described, and its responses to the RF drive power and beam excitation were discussed. Control methods, such as beam loading compensation and feedback control strategies, were proposed based on the TW structure model. Tests were conducted at SwissFEL to validate the model, including the study of the single-bunch-induced load signal and the effects of RF feedback on beam stability. The test results provided preliminary validation of the model's accuracy.

As for the next steps, we will apply the TW structure model to measure and adjust the accelerating voltages and phases of the two bunches that are accelerated within the same RF pulse [15]. This is the standard operation mode of SwissFEL, enabling the simultaneous supply of two beamlines at the full repetition rate (100 Hz). This study is critical for developing a robust method to independently manipulate multiple bunches within a single train – a necessary step for the future upgrade of SwissFEL, which involves the introduction of additional beamlines.